\documentclass[runningheads]{cl2emult}

\usepackage{makeidx}  
\usepackage{graphicx} 
\usepackage{subeqnar} 
\usepackage{multicol} 
\usepackage{cropmark} 
\usepackage{eso}      
\makeindex            

\def\la{\;
\raise0.3ex\hbox{$<$\kern-0.75em\raise-1.1ex\hbox{$\sim$}}\; }
\def\ga{\;
\raise0.3ex\hbox{$>$\kern-0.75em\raise-1.1ex\hbox{$\sim$}}\; }

\begin{document}
\title*{
Molecular hydrogen abundance in the 
dust-free damped Ly-$\alpha$ galaxy
at $z = 3.4$ 
}
\toctitle{
Molecular hydrogen abundance in the 
dust-free damped Ly-$\alpha$ galaxy
at $z = 3.4$ 
}
%
%
\titlerunning{H$_2$ in the $z = 3.4$ DLA galaxy}
%
\author{Sergei~A.~Levshakov\inst{1}
\and Paolo Molaro\inst{2}
\and Miriam Centuri\'{o}n\inst{2}
\and Sandro D'Odorico\inst{3}
\and Piercarlo Bonifacio\inst{2}
\and Giovanni Vladilo\inst{2} }
\authorrunning{Sergei A. Levshakov et al.}
%
%
\institute{Department of Theoretical Astrophysics, 
Ioffe Physico-Technical Institute, 194021 St.~Petersburg, Russia
\and 
Osservatorio Astronomico di Trieste, Via G. B. Tiepolo 11,
34131 Trieste, Italy
\and
European Southern Observatory, 85748 Garching bei M\"{u}nchen, Germany
     }

\maketitle              

\begin{abstract}
New results from the search for H$_2$ absorption
in the damped Ly$\alpha$ galaxy at redshift $z = 3.4$
toward QSO 0000--2620 ($z_{\rm em} = 4.1$) are reported.
The high-resolution ($\lambda/\Delta\lambda = 48,000$)
spectra of Q0000--2620
were obtained using the Ultraviolet - Visual Echelle
Spectrograph (UVES) on the 8.2m {\itshape ESO} Kueyen telescope.
The ortho-H$_2$ column density
is found to be $N(J=1) = (5.55 \pm 1.35)\times10^{13}$ cm$^{-2}$ 
($2\sigma$ C.L.).
The combination of $N(J=1)$ with the limits
available for other low rotational levels restricts
the excitation temperature $T_{\rm ex}$ in the range $(290 - 540)$~K.   
This gives the total H$_2$ column density of 
$N({\rm H}_2) = (8.75 \pm 1.25)\times10^{13}$ cm$^{-2}$ and
the corresponding fraction of hydrogen atoms bound in molecules
of $f({\rm H}_2) = (6.8 \pm 2.0)\times10^{-8}$.
\end{abstract}

\section{Introduction}

It has long been recognized that H$_2$ (and HD) molecules
play a central role in the formation of gas condensations
in the post-recombination era since they provide the cooling
necessary for the collapse on all scales of the first objects.
In the primordial gas at redshift $z \la 50$, 
the fractional abundance of H$_2$ is
calculated to be $f({\rm H}_2) = 10^{-5} - 10^{-6}$ 
(e.g. \cite{ref1.1}).
The H$_2$ absorption lines from the Lyman and Werner bands
may be observable at lower redshift $z \sim 2 - 4$
in QSO absorption systems with high 
neutral hydrogen column densities, $N({\rm HI}) = 10^{21} - 10^{22}$
cm$^{-2}$ (so-called Damped Lyman $\alpha$ systems, DLA).
It has been suggested that these systems are most closely
related to the progenitors of normal galaxies~\cite{ref1.2}. 
So far, absorption from H$_2$ has been detected in 
a few DLA systems which show, in general, a small amount of 
molecular gas~\cite{ref1.2a}. 
In contrast to DLAs, observations in our Galaxy reveal
the presence of H$_2$ in nearly all lines of sight in the disk
and halo~\cite{ref1.3}.
It was also found that the ISM diffuse clouds with the neutral hydrogen
column densities
$N({\rm HI}) \ga 3\times10^{20}$ cm$^{-2}$
show 
$f({\rm H}_2) \ga 10^{-5}$.

\begin{figure}
\vspace{-1.5cm}
\centering
\includegraphics[width=1.0\textwidth]{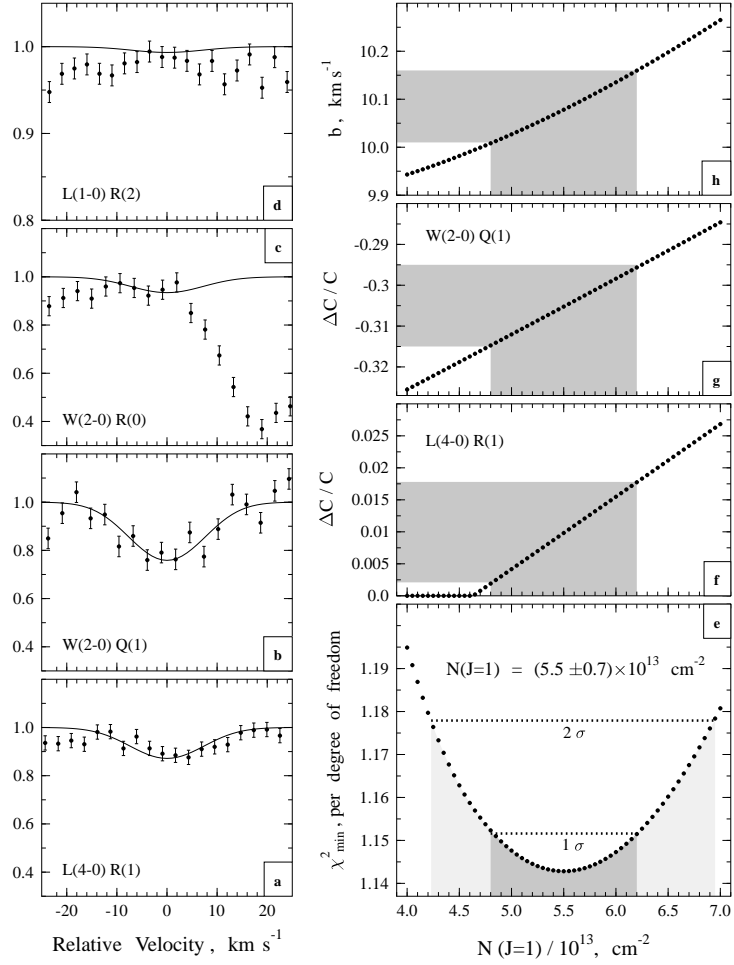}
\vspace{-1.0cm}
\caption[]{ Best fit to H$_2$ lines associated with the $z = 3.4$ DLA
system toward Q0000--2620. ({\bf a}, {\bf b}, {\bf c}, {\bf d}) 
Simultaneous fit
to the Lyman L(4-0)R(1), L(1-0)R(2) 
and Werner W(2-0)Q(1), W(2-0)R(0) lines
with $N(J=0) = 8.4\times10^{12}$ cm$^{-2}$ ($1\sigma$ upper limit),
$N(J=1) = (5.5 \pm 0.7)\times10^{13}$ cm$^{-2}$, 
$N(J=2) = 1.2\times10^{13}$ cm$^{-2}$ ($1\sigma$ upper limit),
and $b \simeq 10.07$
km~s$^{-1}$. The velocities shown are related to $z = 3.390127$.
({\bf e}) Confidence regions for the ortho-H$_2$ column density
calculated from the simultaneous fit of the L(4-0)R(1) + W(2-0)Q(1) lines
and all low-ion lines located redward the Ly$\alpha$ emission.
({\bf f}, {\bf g}) The corresponding deviations of the local continuum
level. ({\bf h}) The corresponding Doppler parameter variations.
The grey areas restrict the $\Delta C / C$ and $b$ values at $1\sigma$
level in accordance with panel {\bf e}
}
\label{eps1}
\end{figure}

Absorption from H$_2$ in the $z = 3.4$ DLA system toward Q0000--2620
was not detected in the 1 \AA\, resolution spectrum obtained with the
Multiple Mirror Telescope (MMT) and only an upper limit of
$f({\rm H}_2) < 3\times10^{-6}$ has been reported previously~\cite{ref1.4}.  
New observations with the UVES/VLT, having approximately 11 times higher
resolution, revealed H$_2$ absorption in this DLA.
Exactly at the expected position of the L(4-0)R(1) line where the MMT 
gave the limit $W_{\rm rest} < 114$ m\AA\, (3$\sigma$), 
an absorption line with 
the equivalent width $W_{\rm rest} \simeq 6$ m\AA\, 
was detected~\cite{ref1.5}. 
The chance identification of this line in the Ly$\alpha$ forest
at $z \sim 3$ has a probability $\la 10^{-3}$ (cf.~\cite{ref1.5a}).

Below we report on new identifications of H$_2$
lines in this DLA system and give improved values of molecular
hydrogen column density and excitation temperature.

\section{Data analysis and results}

Spectroscopic observations of Q0000--2620 are described in detail 
in~\cite{ref1.6}. 
The spectrum was obtained with the rms uncertainty
of the wavelength calibration $\delta \lambda \leq 0.6$ km~s$^{-1}$,
the velocity resolution of FWHM $\simeq 6$ km~s$^{-1}$
(the corresponding bin size equals 2.4 km~s$^{-1}$)
and the signal-to-noise
ratio of S/N $\simeq 40$ (per pixel)
in the range $\lambda\lambda = 4605 - 4615$ \AA\,
allowing us to detect the L(4-0)R(1) line 
($\lambda_0 = 1049.9596$ \AA\, and the oscillator strength
$f = 0.016$ are from~\cite{ref1.7}) at the
expected position, $\lambda_{\rm obs} = 4609.4$ \AA.
The line shows excellent redshift agreement with the low-ion lines
and the same Doppler parameter, $b \simeq 10$ km~s$^{-1}$.
The corresponding H$_2$ lines from the Lyman and Werner bands are
badly blended with Ly$\alpha$ forest absorption, but the L(4-0)R(1) line
is relatively clean of contamination from forest absorption.

The analysis of absorption lines occurring in the Ly$\alpha$ forest may
have large intrinsic errors due to uncertainties in
the local continuum level determinations.
Therefore to control solutions, theoretical L(4-0)R(1) profiles were fitted
to the observed intensities simultaneously with 
the fitting of all metal lines 
located redward the QSO Ly$\alpha$ emission (ZnII$\lambda2026$,
CrII$\lambda2062$, CrII$\lambda2056$, FeII$\lambda1611$,
SiII$\lambda1808$, and NiII$\lambda1709$), assuming that $b$ is the same
for H$_2$ and low-ion lines and leaving all the elemental column 
densities as well as
the local continuum deviations ($\Delta C / C$) 
around the L(4-0)R(1) line free
to vary (the computational procedure is described in~\cite{ref1.8}).

Analyzing the obtained UVES spectrum,
we have identified  a new line W(2-0)Q(1) from  
the Werner band, and have set a more stringent upper limit to the
para-H$_2$ abundance through the W(2-0)R(0) line. The fits are
shown in Fig.~1 ({\bf a}, {\bf b}, {\bf c}, {\bf d}) 
by solid lines superimposed on the observed
(re-normalized) intensities, which are marked by dots with 1$\sigma$ error
bars. A simultaneous treatment of the L(4-0)R(1) and W(2-0)Q(1) lines yields
the ortho-H$_2$ column density of $N(J=1) = (5.5 \pm 0.7)\times10^{13}$
cm$^{-2}$. 
Right column of Fig.~1 shows 
the confidence levels for $N(J=1)$ (panel {\bf e}),
the corresponding variations for $\Delta C / C$ 
(panels {\bf f} and {\bf g}), and the Doppler broadening (panel {\bf h}).
From panels {\bf f} and {\bf g} it is seen that the local continuum 
is shifted with respect to the general
continuum level 
(determined over the whole region of the Q0000--2620 spectrum)
to about +~1.0\% for the L(4-0)R(1) line and to --~30.5\%
for the W(2-0)Q(1) line, respectively. 
The shifts at the position of the W(2-0)R(0) line (panel {\bf c}) 
and at the position of the L(1-0)R(2) line (panel {\bf d})
are found to be negligible.
The combination of the $N(J=1)$ value with the limits
$N(J=0) \la 8.4\times10^{12}$ cm$^{-2}$ and 
$N(J=2) \la 1.2\times10^{13}$ cm$^{-2}$  
restricts the excitation temperature $T_{\rm ex}$ in the 
range $(290 - 540)$~K.   
This gives the total H$_2$ column density of 
$N({\rm H}_2) = (8.75 \pm 1.25)\times10^{13}$ cm$^{-2}$ and
the corresponding fraction of hydrogen atoms bound in molecules
of $f({\rm H}_2) = (6.8 \pm 2.0)\times10^{-8} $\,\,\cite{ref1.9}.

The obtained result has an interesting connection to a local DLA system --
the metal-deficient (Z/Z$_\odot \sim 1/50$) starburst galaxy I~Zw~18
with $N({\rm HI}) \simeq 3\times10^{21}$ cm$^{-2}$ where recent FUSE
observations set a limit $f({\rm H}_2) \ll 10^{-6}$\,\,\cite{ref1.10}.

\vspace{0.5cm}

{\it Acknowledgments}.  
The work of S.A.L. is supported in part 
by the Deutsche Forschungsgemeinschaft
and by the RFBR grant No.~00-02-16007.

\clearpage
\addcontentsline{toc}{section}{Index}
\flushbottom
\printindex

\end{document}